\documentclass[12pt, a4paper]{article}
\usepackage[utf8]{inputenc}
\usepackage{amsmath, amssymb}
\usepackage{graphicx}
\usepackage{geometry}
\usepackage{booktabs}
\usepackage{cite}
\usepackage{hyperref}
\usepackage{float} 
\title{SuperEM: A Sub-meV Threshold Detector Architecture for Cosmic Neutrino Background and Dark Matter Detection}
\author{
    Zhenjie Li$^1$, Xilei Sun$^{1,*}$, Xiaoshan Jiang$^1$\\[1ex]
    $^1$\textit{Institute of High Energy Physics, Chinese Academy of Sciences,}\\
    \textit{Beijing 100049, China}\\[1ex]
    $^*$Contact author: \texttt{sunxl@ihep.ac.cn}
}
\date{}

\begin{document}

\maketitle

\begin{abstract}
Expanding the operational boundaries of radiation detection is imperative for contemporary particle physics, astrophysics, and cosmology. At this frontier, the direct detection of the Cosmic Neutrino Background (C$\nu$B), the determination of the absolute neutrino mass scale, and the search for sub-GeV Light Dark Matter (LDM) necessitate detector architectures capable of sub-millielectronvolt (sub-meV) energy thresholds, exceptional absolute energy resolution, fast time response, and massive scalability. Current technologies confront an intrinsic limit---the ``impossible triangle''---wherein optimizing for sub-meV thresholds inherently compromises either macroscopic timing response or spatial scalability. Here, we introduce the Superconductor-Coupled Semiconductor Electron-Multiplying (SuperEM) detector, a fundamentally novel structural paradigm designed to bypass this limitation. The architecture couples the ultra-low energy threshold of a superconducting absorber with the intrinsic high-gain digitization of a strongly biased, high-density semiconductor P-N junction. Incident energy yields a proliferation of non-equilibrium quasiparticles, which are subsequently extracted via quantum tunneling across an ultra-thin Atomic Layer Deposition (ALD) insulating barrier. Building upon our prior empirical validation of deep-cryogenic avalanche mechanics, this manuscript establishes the fundamental theoretical feasibility and structural foundation of the complete device. By analytically demonstrating that a Pixelated Geiger-Counting Mode is mandatory to bypass the proportional variance penalty, we establish that the architecture achieves a rigorous absolute energy resolution of $\sim 43.8$~meV for a 1~eV deposition, decisively encroaching upon the limits required for neutrino mass ordering differentiation. Furthermore, signal transport simulations confirm that an undoped interface coupled with a strong drift field enables highly efficient, nanosecond-scale transient electron drift, resolving completely within 35~ns. The SuperEM architecture thus constitutes a scalable, high-resolution, and fast time-response framework for next-generation C$\nu$B and LDM observatories.
\end{abstract}

\section{Introduction}
The advancement of human knowledge has always been inextricably linked to the tools we build to observe it. From the telescopes that unveiled the macroscopic architecture of the cosmos to the scintillation detectors that exposed the microscopic dynamics of radioactivity, every major leap in our understanding of nature has required a corresponding breakthrough in perception. Currently, the extreme low-energy frontier—specifically the sub-millielectronvolt (sub-meV) regime—stands as the definitive boundary of our observational capacity. Crossing this threshold to effectively detect such minuscule energy depositions will open an unprecedented window for physics, allowing us to interact with the universe in ways previously thought impossible.

At the forefront of this new observational epoch is the pursuit of the Cosmic Neutrino Background (C$\nu$B), a cosmological relic consisting of neutrinos that decoupled from baryonic matter approximately one second after the Big Bang \cite{weinberg2008}. According to standard cosmological models, these relic neutrinos presently permeate the universe with an average kinetic energy of roughly $10^{-4}$~eV, an equivalent temperature of 1.95~K, and a density approximating 300 particles per cubic centimeter \cite{weinberg2008}.

The established detection strategy relies on the neutrino capture reaction by beta-unstable nuclei, specifically the reaction $\nu_e+^3\text{H}\rightarrow^3\text{He}+e^-$ \cite{weinberg1962}. This weak-interaction process possesses no energy threshold and generates a signal characterized by a monoenergetic electron located just beyond the endpoint energy of the standard tritium beta-decay spectrum ($Q_\beta\approx 18.6$~keV) \cite{weinberg1962}. To extract this infinitesimal signal peak from the overwhelmingly intense continuous beta-decay background, observatories require an absolute detector energy resolution superior to 50~meV, alongside an ability to process extreme background counting rates exceeding $10^5$~Hz/g without incurring spectral distortion \cite{baracchini2018}.

Alternatively, newly advanced theoretical frameworks propose detecting the C$\nu$B through the coherent parametric fluorescence of relic neutrinos interacting within cold atomic or molecular systems \cite{huang2026}. In this process ($\nu_i+M\rightarrow\nu_j+\gamma_s+M$), a heavy relic neutrino scatters coherently with molecular dipoles, transitioning into a lighter neutrino and emitting a signal photon ($\gamma_s$). For a normal mass ordering scenario, this transition yields a far-infrared photon with an energy of approximately 25~meV \cite{huang2026}. Practical application of this method demands sensors capable of single-photon detection at $\mathcal{O}(10)$~meV energies, scaled over macroscopic target geometries (e.g., 5~cm$^3$) \cite{huang2026}.

Similarly, the laboratory determination of the absolute neutrino mass scale remains a critical unresolved parameter in the Standard Model. Because current neutrino oscillation data physically constrains the minimum effective electron antineutrino mass for the inverted hierarchy to approximately 50~meV, achieving an absolute instrumental energy resolution strictly below this scale (typically $\leq 40$~meV) is a prerequisite to kinematically distinguish the normal versus inverted mass ordering via tritium beta-decay endpoint measurements \cite{aker2019, ashtari2017}. Additionally, Coherent Elastic Neutrino-Nucleus Scattering (CE$\nu$NS) experiments also require detector thresholds as low as the meV level to achieve detector miniaturization \cite{freedman1974, akimov2017}.

Beyond neutrino physics, the search for Light Dark Matter (LDM) in the sub-GeV mass range represents another critical observational frontier. Because these ultra-light candidates possess masses far below the traditional Weakly Interacting Massive Particle (WIMP) scale, their scattering kinematics transfer infinitesimal amounts of kinetic energy to the detector medium. Probing this largely uncharted parameter space therefore inherently necessitates detector architectures with strictly sub-meV sensitivities \cite{hochberg2016}. 

Parallel to these fundamental particle physics searches, the mitigation of ``quasiparticle poisoning'' in superconducting quantum computers demands real-time detectors with nanosecond-level response times and comparable sub-meV sensitivities \cite{li2025}. Furthermore, the quest to detect primordial gravitational waves via the B-mode polarization of the Cosmic Microwave Background (CMB) demands the deployment of massively scaled, ultra-sensitive large-area microwave detector arrays capable of achieving unprecedented performance and high multiplexing factors across tens of thousands of channels \cite{abazajian2019}. This imperative is distinctly highlighted by major observational initiatives such as the Ali CMB Polarization Telescope (AliCPT) in Tibet, China, which is designed to probe primordial gravitational waves from the northern hemisphere by deploying vast, highly sensitive arrays of transition-edge sensors \cite{li2017}.

Historically, the pursuit of ultimate sensitivity in neutrino and dark matter observatories has predominantly relied upon modifying the properties of the detection medium and exponentially increasing the target mass, with modern detectors reaching scales of tens of thousands of tons. While increasing the macroscopic scale of the detection medium remains crucial for capturing extremely rare interactions, this ``medium-centric'' approach cannot fundamentally bypass intrinsic energy-threshold barriers. To access the sub-meV regime required for C$\nu$B and LDM detection, the field must transition from ``medium modification'' to ``underlying architectural innovation.''

Achieving an ultra-low sub-meV threshold through innovation in the detector’s fundamental architecture will drastically redefine physical sensitivity. Rather than relying on the brute-force scaling of macroscopic target masses, the next epoch of rare-event searches necessitates detectors capable of directly extracting and amplifying the intrinsic fluctuations of microscopic quantum states. Existing low-temperature quantum sensing technologies represent a major leap in sensitivity, yet they are restricted by fundamental physical limitations that prevent the simultaneous satisfaction of these requirements.

Cryogenic detectors broadly exploit ultra-low superconducting energy gaps to achieve high resolution. For instance, Superconducting Tunnel Junction (STJ) sensors utilize five-layer thin-film devices to break Cooper pairs in materials like Tantalum ($\Delta_{\text{Ta}}=0.7$~meV), delivering an energy resolution approximately 30 times superior to conventional silicon or germanium architectures \cite{fretwell2020}. This high resolution capacity has enabled breakthrough experiments in low-energy precision subatomic physics, including the first direct measurement of the $^7$Be L/K orbital electron capture ratio in $^7$Be via nuclear-recoil spectroscopy \cite{fretwell2020}. However, as summarized alongside other mainstream detection frameworks in Table 1, existing physical architectures remain bounded by key technical bottlenecks.

\begin{table}[H]
\centering
\caption{The performance landscape of current mainstream detection architectures.}
\small
\begin{tabular}{@{} p{3.5cm} p{2.5cm} p{2cm} p{3.5cm} p{3.5cm} @{}}
\toprule
\textbf{Detector Architecture} & \textbf{Representative Project} & \textbf{Energy Threshold} & \textbf{Reading Complexity \& Scalability} & \textbf{Temporal Response} \\
\midrule
High-Purity Germanium (HPGe) & CONUS \cite{bonet2021} & $\sim$100~eV & Moderate & Nanoseconds \\
Skipper-CCD & SENSEI \cite{abramoff2019}, Oscura \cite{aguilar2022, cervantes2023} & $\sim$1~eV & High (Precision multi-sampling) & Milliseconds to Hours \\
Transition Edge Sensor (TES) & PTOLEMY \cite{baracchini2018}, AliCPT \cite{li2017} & $\sim$100~meV & Extremely High (SQUID multiplexing) & Microseconds to Milliseconds \\
Kinetic Inductance Detector (KID) & CCAT-prime \cite{choi2020} & $\sim$1~eV & Moderate (Frequency multiplexing) & Microseconds \\
Superconducting Nanowire (SNSPD) & QROCODILE \cite{natarajan2012} & $\sim$100~meV & Low & Picoseconds \\
Superconducting Tunnel Junction (STJ) & $^7$Be Nuclear Recoil \cite{fretwell2020} & $\sim$10~eV & Moderate to High & Nanoseconds \\
\bottomrule
\end{tabular}
\end{table}

This intersecting technological impasse forms the ``impossible triangle'' of rare-event detection: current physical architectures cannot simultaneously achieve sub-meV energy thresholds, nanosecond-scale temporal responses, and the massive parallel scalability required for large-scale experiments. Although architectures such as STJs exhibit remarkable high-rate performance for localized radioactive isotope recoils \cite{fretwell2020}, their lack of internal amplification of electrons and reliance on preamplifier readout signals limit their application in ultra-low threshold detection.

To dismantle this triad of constraints, we propose the Superconductor-Coupled Semiconductor Electron-Multiplying (SuperEM) detector. By isolating the microscopic energy sensing phase across a quantum tunneling barrier from the macroscopic avalanche process, the SuperEM architecture successfully couples the ultra-low energy threshold of a superconductor with the intrinsic, high-gain amplification of a semiconductor. Conceptually, this architecture operates analogously to a photomultiplier tube (PMT): the superconducting film serves as the equivalent of a photocathode, where the breaking of Cooper pairs mirrors the photoelectric effect. Subsequent electron multiplication is then utilized for massive signal amplification, facilitating robust macroscopic electronic readout. Furthermore, by drawing a direct parallel to the established grid-based architecture of Silicon Photomultipliers (SiPMs), the inherent large-area spatial scalability of the SuperEM detector becomes readily apparent.

To realize this transformative technology, our group has initiated a comprehensive, multi-phase research program. In our first phase \cite{gao2025}, we addressed the most critical operational uncertainty of the semiconductor backend by experimentally proving that silicon avalanche multiplication persists under severe carrier freeze-out at 10~mK.

Building directly upon that empirical milestone, this manuscript serves as the definitive theoretical and architectural foundation for the fully integrated SuperEM detector. While the basic physical mechanisms of the components have been validated, the realization of a functional device requires a highly specific topology. Herein, we systematically resolve the remaining physical uncertainties regarding quantum tunneling, interface transport mechanics, and statistical variance through rigorous analytical derivation and TCAD signal simulation. By demonstrating the necessity of a pixelated Geiger-mode topology, we clear the theoretical pathway for the physical realization of the complete device.

\section{Architectural Framework}
As demonstrated in Figure 1(a), the SuperEM detector relies on a highly engineered heterostructure consisting of three distinct physical layers: a Superconductor, an Insulator, and a PN junction semiconductor layer (S-I-P-N). It should also be noted that either a P-type or N-type substrate can be used for the PN junction; this paper uses a P-type substrate as an example. The core conceptual originality resides in seamlessly coupling the ultra-low energy threshold of a superconductor with the intrinsic, high-gain amplification of a semiconductor. Figure 1(b) shows the one-dimensional energy band alignment across the heterojunction at 10~mK. To further elaborate, the key to quasiparticle tunneling into the semiconductor is that the conduction band minimum ($E_c$) of the semiconductor is lower in energy than the non-equilibrium quasiparticle states within the superconductor.

\textbf{Superconducting Absorber (S-Layer):} The primary detection medium is a thin film of a conventional low-critical-temperature ($T_c$) superconductor, for example, Aluminum (Al). Incident particles deposit kinetic energy into the lattice, breaking ground-state Cooper pairs into non-equilibrium quasiparticles \cite{bardeen1957}. Because the binding energy of a Cooper pair in aluminum is $\sim$0.35~meV \cite{bardeen1957}, a 1~eV energy deposition generates thousands of primary signal carriers, effectively overriding the Poisson limits inherent to semiconductors.

\textbf{Quantum Tunneling Barrier (I-Layer):} An ultra-thin insulating layer—such as an aluminum oxide (Al$_2$O$_3$) insulating film deposited via ALD—acts as a selective quantum filter. It permits the ``cold injection'' of non-equilibrium quasiparticles into the semiconductor via wave-function tunneling. More importantly, applying a bias voltage across this isolation layer dynamically regulates the tunneling probability, functioning analogously to a gating mechanism; modulating this bias voltage enables precise control over the sensor’s active state (on/off) and its energy threshold, while simultaneously blocking the reverse leakage of high-energy electrons from the macroscopic avalanche region.

\textbf{Charge Multiplication Region (P-N Junction):} Deeply cooled to 10~mK and subjected to a high reverse bias, primary electrons undergo impact ionization upon entering the high-field depletion region \cite{sze2006}. The semiconductor acts as a built-in pre-amplifier with immense gain, circumventing the need for highly complex SQUID arrays and ensuring that the macroscopic signal readout remains fully compatible with conventional nuclear electronics.

\begin{figure}[H]
\centering
\includegraphics[width=0.9\textwidth]{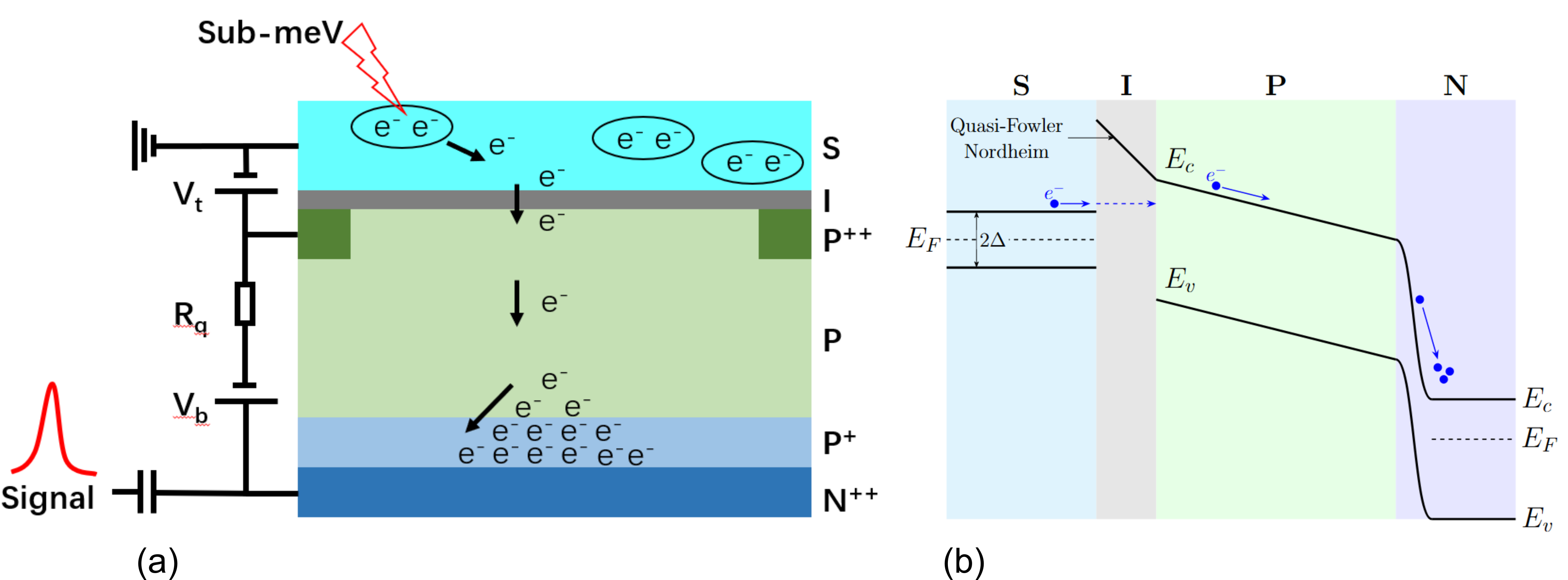}
\caption{Conceptual architecture and energy band diagram of the SuperEM detector. (a) Cross-sectional schematic of the S-I-P-N heterostructure. An incident rare-event particle deposits energy into the superconducting aluminum absorber (S-layer), breaking Cooper pairs into non-equilibrium quasiparticles. These carriers quantum-tunnel across the ultra-thin ALD insulator (I-layer) into the neutral P-type semiconductor (P-layer), which is laterally bounded by a highly doped P$^{++}$ contact electrode. Driven by an external bias voltage, the injected electrons are swept into the P-N junction toward the highly doped N$^{++}$ anode bounding the entire bottom surface, initiating a macroscopic avalanche with a high gain. (b) One-dimensional energy band alignment across the heterojunction at 10~mK. The diagram depicts the 2$\Delta$ superconducting gap, the Quasi-Fowler-Nordheim triangularized tunneling barrier shaped by a strong internal bias, and the externally applied electric field that induces saturated electron drift within the semiconductor.}
\end{figure}

To address diverse experimental requirements, the fundamental S-I-P-N physics can be conceptualized via a baseline \textbf{Monolithic Proportional Mode}: The P-N junction is biased slightly below its breakdown voltage ($V_{\text{bias}} < V_{\text{bd}}$), yielding moderate internal gain ($M \approx 10^2$ to $10^3$). Total output charge remains strictly proportional to the initial quasiparticle generation.

However, to achieve the meV resolution and massive macroscopic scalability required by modern rare-event observatories, the architecture demands a \textbf{Pixelated Geiger-Counting Mode}: The substrate is lithographically segmented into a high-density array of micro-pixels equipped with quenching resistors, and biased beyond the breakdown voltage. A single tunneled electron triggers a saturated discharge within a localized pixel, entirely eradicating analog continuous avalanche multiplication noise \cite{acerbi2019}.

\section{Feasibility of Physical Principles}
The theoretical viability of the SuperEM detector rests upon the sequential execution of Cooper pair dissociation, quantum tunneling, electron transit, and macroscopic avalanche multiplication.

\subsection{Superconducting Absorption and Cooper Pair Dynamics}
In standard semiconductors, the average energy ($\epsilon$) required to generate an electron-hole pair is approximately 3.6~eV for silicon \cite{knoll2010}. In an aluminum superconductor operating significantly below its critical temperature, the half-gap is $\Delta\approx 0.35$~meV \cite{bardeen1957}. The irreversible partition of energy between the measurable quasiparticle system and irrecoverable lattice heat dictates the effective average energy ($\epsilon_{\text{eff}}$) required to produce a single stable quasiparticle \cite{kurakado1982}:
\begin{equation}
\epsilon_{\text{eff}} = 1.7\Delta \approx 0.595\text{ meV}
\end{equation}
where 1.7 is the phenomenological constant for elemental superconductors defining the ratio of usable energy to lattice loss—specifically accounting for the irreversible energy lost to the creation of sub-gap phonons during the initial relaxation cascade—and $\Delta$ is the superconducting half-gap. This sub-meV effective excitation energy yields thousands of times more signal carriers than a semiconductor for an equivalent deposition. The intrinsic energy resolution (FWHM) due solely to Cooper pair generation fluctuations represents the absolute physical floor of the detector’s capability \cite{knoll2010}:
\begin{equation}
R_{\text{int}} = 2.355\sqrt{F\epsilon_{\text{eff}}E_{\text{dep}}}
\end{equation}
where $E_{\text{dep}}$ is the total deposited kinetic energy, $F\approx 0.2$ is the Fano factor for elemental superconductors \cite{kurakado1982}, $\epsilon_{\text{eff}}$ is the effective pair-creation energy, and the prefactor 2.355 converts the statistical standard deviation ($\sigma$) to the Full Width at Half Maximum (FWHM). At the crucial 1~eV energy scale, the intrinsic physical limit of the aluminum absorber evaluates to 25.7~meV, establishing sufficient physical margin for the 40~meV instrument target required to resolve the inverted neutrino mass ordering \cite{aker2019}. Furthermore, to achieve the extreme sensitivity required to probe the normal-order neutrino mass hierarchy, the architectural medium can be substituted with a superconductor possessing a narrower Cooper pair half-gap, such as Hafnium ($\Delta\approx 0.02$~meV) \cite{friedrich2026}. Under this configuration, the effective pair-creation energy drops to 0.034~meV. For a 100~meV energy deposition, the theoretical intrinsic energy resolution ($R_{\text{int}}$) strictly evaluates to approximately 1.94~meV. This sub-3~meV resolution capacity firmly satisfies the rigorous instrumental prerequisites for future normal-hierarchy absolute mass measurement experiments.

\subsection{Quantum Tunneling Dynamics}
Transverse transport of non-equilibrium quasiparticles across the ultra-thin insulating layer (Al$_2$O$_3$) at 10~mK is strictly governed by quantum mechanical wave-function tunneling. The baseline microscopic tunneling probability ($D$) is analytically evaluated using the Wentzel-Kramers-Brillouin (WKB) approximation \cite{simmons1963}:
\begin{equation}
D(E_x) = \exp\left(-\frac{4\pi}{h}\int_{x_1}^{x_2}\sqrt{2m^*[V(x)-E_x]}\,dx\right)
\end{equation}
where $h$ is Planck’s constant, $x_1$ and $x_2$ are the classical turning points defining the physical boundaries of the tunneling barrier, $m^*$ is the effective electron mass within the Al$_2$O$_3$ lattice (typically $\sim 0.4m_e$ for ALD-deposited films \cite{yeo2003}), $V(x)$ is the position-dependent potential energy profile of the barrier, and $E_x$ is the transverse kinetic energy of the incident quasiparticle.

When an external reverse bias ($V_b$) drops directly across the ultra-thin insulator of thickness $d$, it induces a massive internal electric field ($E=V_b/d$). According to the generalized Simmons model \cite{simmons1963}, this strong bias severely triangularizes the barrier profile $V(x)$, triggering Quasi-Fowler-Nordheim tunneling. This process compresses the effective barrier width, substantially increasing the tunneling probability.

In this tunneling framework, the transverse kinetic energy $E_x$ of the incident quasiparticles is evaluated relative to the bottom of the conduction band and is approximated by the Fermi energy of the aluminum absorber ($E_F \approx 11.7$~eV) \cite{ashcroft1976}. The sub-meV non-equilibrium excitation energy ($\epsilon_{eff} \approx 0.595$~meV) contributes a negligible perturbation to the tunneling dynamics. The position-dependent potential profile $V(x)$ is defined by a baseline, zero-field Al--Al$_2$O$_3$ effective interface barrier height of $\Phi_B = 1.5$~eV \cite{wilt2017} relative to the Fermi level. Because this barrier height perfectly matches the applied internal bias of $V_b = 1.5$~V, the barrier completely loses its trapezoidal geometry and undergoes severe triangularization into a pure Quasi-Fowler-Nordheim profile under the intense local electric field. Inputting these boundary conditions along with an effective electron mass of $m^* = 0.4m_e$ and an insulating layer thickness of $d = 2.0$~nm into Formula (3) yields a precise, static microscopic tunneling probability of $D \approx 2.53 \times 10^{-5}$.

To precisely quantify the final device efficiency, the system must be modeled as a kinetic competition between the active tunneling rate ($\Gamma_{\text{tun}}$) and the intrinsic quasiparticle recombination loss rate ($\Gamma_{\text{loss}}$) within the aluminum absorber. This determines the macroscopic extraction efficiency ($\eta_t$):
\begin{equation}
\eta_t = \frac{\Gamma_{\text{tun}}}{\Gamma_{\text{tun}}+\Gamma_{\text{loss}}}
\end{equation}
The tunneling rate is defined as $\Gamma_{\text{tun}}=\nu_0D$, where the attempt-to-escape frequency ($\nu_0$) is governed by the Fermi velocity of the charge carriers ($v_F$) and the physical thickness of the superconducting aluminum film ($d_{\text{Al}}$). Specifically, the attempt-to-escape frequency is defined as $\nu_0 = v_F / 4d_{\text{Al}}$ \cite{krauss1989}. Given the Fermi velocity of aluminum ($v_F \approx 2.02 \times 10^{15}\text{ nm/s}$ \cite{ashcroft1976}), this kinetic parameter is strictly geometry-dependent. In high-purity aluminum at 10~mK, thermal recombination is drastically suppressed by carrier freeze-out, resulting in an exceptionally slow intrinsic loss rate ($\Gamma_{\text{loss}}$) on the order of $10^3$ to $10^4\text{ s}^{-1}$ \cite{kaplan1976}.

Consequently, the macroscopic extraction efficiency ($\eta_t$) of quasi-particles quantum tunneling into the semiconductor is primarily determined by three highly controllable parameters: the thickness of the superconducting aluminum film ($d_{\text{Al}}$), the thickness of the insulating layer ($d$), and the internal bias voltage ($V_b$).

Using this kinetic framework, we calculated a two-dimensional thermodynamic surface plot of $\eta_t$ as a function of the insulating layer thickness and bias voltage. By assuming a typical optimal aluminum film thickness of 50~nm (yielding an attempt-to-escape frequency of $\nu_0\approx 10^{13}$~Hz), the dynamic extraction efficiency can be mapped. As demonstrated in Figure 2(a), operating within the safe dielectric breakdown limit ($E\leq 10^7\text{ V/cm}$) \cite{groner2002}, the macroscopic extraction efficiency can be robustly tuned from 0\% to nearly 100\%. This dynamic modulation allows the bias voltage to act as a highly effective gating mechanism. Specifically, for an insulating layer thickness of $d=2.0$~nm and an applied bias of $V_b=1.5$~V, the resulting transmission probability heavily outpaces thermal loss, allowing $\eta_t$ to reach an exceptional efficiency of over 99\%.

Furthermore, we evaluated the direct impact of the superconducting aluminum film thickness ($d_{\text{Al}}$) on $\eta_t$ while maintaining the optimized barrier parameters ($d=2.0$~nm, $V_b=1.5$~V). Because thicker films increase the quasiparticle transit distance and subsequently lower the attempt-to-escape frequency ($\nu_0$), the macroscopic extraction efficiency strictly increases as the aluminum film thickness decreases. As quantitatively illustrated in the logarithmic plot of Figure 2(b), the architecture is highly forgiving within the ultra-thin film regime, achieving a near-perfect extraction efficiency of 98.1\% at $d_{\text{Al}} = 100$~nm. However, as the absorber thickness scales into the micrometer regime, the prolonged transit time allows the intrinsic thermal recombination loss ($\Gamma_{\text{loss}}$) to become increasingly competitive against the tunneling rate. Consequently, the extraction efficiency systematically degrades to 83.6\% at a thickness of 1~$\mu$m ($10^3$~nm) and drops precipitously to 33.8\% at 10~$\mu$m ($10^4$~nm). This strong geometric dependence dictates that the superconducting absorber must be fabricated strictly as an ultra-thin film (ideally $\le 100$~nm) to preserve the integrity of the sub-meV signal prior to avalanche multiplication.

The mean tunneling time ($\tau_{\text{tun}}$) is defined as the inverse of the active tunneling rate ($1/\nu_0D$). Substituting the previously established attempt-to-escape frequency ($\nu_0 = v_F / 4d_{\text{Al}}$) demonstrates that the tunneling time scales linearly with the absorber’s geometry \cite{krauss1989}:
\begin{equation}
\tau_{\text{tun}} = \frac{4d_{\text{Al}}}{v_FD}
\end{equation}
Under the optimized barrier conditions ($D\approx 2.53\times 10^{-5}$), this relationship simplifies to $\tau_{\text{tun}}\approx 0.0783\times d_{\text{Al}}$ (in nanoseconds). Consequently, when the aluminum film thickness is restricted to less than 100~nm, the quantum tunneling process resolves in under 7.8~ns. Because this transit is overwhelmingly faster than the intrinsic thermal recombination lifetime, the macroscopic extraction efficiency ($\eta_t$) consistently exceeds 98\%, ensuring near-perfect signal conservation prior to avalanche multiplication. Crucially, this nanosecond-scale kinetic extraction constitutes the fundamental physical basis for the rapid temporal response of the SuperEM architecture, allowing the detector to natively bypass the thermal relaxation bottlenecks and macroscopic signal pile-up that plague traditional cryogenic sensors.

It is critical to address the impact of non-ideal barrier conditions, specifically the potential for Trap-Assisted Tunneling (TAT) across the Al$_2$O$_3$ insulating layer. In practical heterostructures, localized defect states at the insulator-semiconductor interface can act as intermediate energy stepping stones, theoretically allowing ``dark'' electrons to cross the barrier and degrade the signal-to-noise ratio. Under standard thermodynamic conditions, TAT is predominantly a thermally activated process. However, within the deep-cryogenic 10~mK environment of the SuperEM detector, phonon-mediated thermal excitation is completely frozen out, aggressively suppressing the thermally activated components of TAT. Despite this profound environmental advantage, elastic (temperature-independent) TAT can still be driven by the strong internal bias ($V_b$) if the spatial density of structural defects is high. Therefore, while the 10~mK operating temperature natively neutralizes thermal dark currents, preventing field-driven elastic TAT from polluting the sub-meV signal strictly relies on achieving a pristine, defect-free atomic layer—a fundamental fabrication imperative that will be detailed in Section 5.

\begin{figure}[H]
\centering
\includegraphics[width=0.9\textwidth]{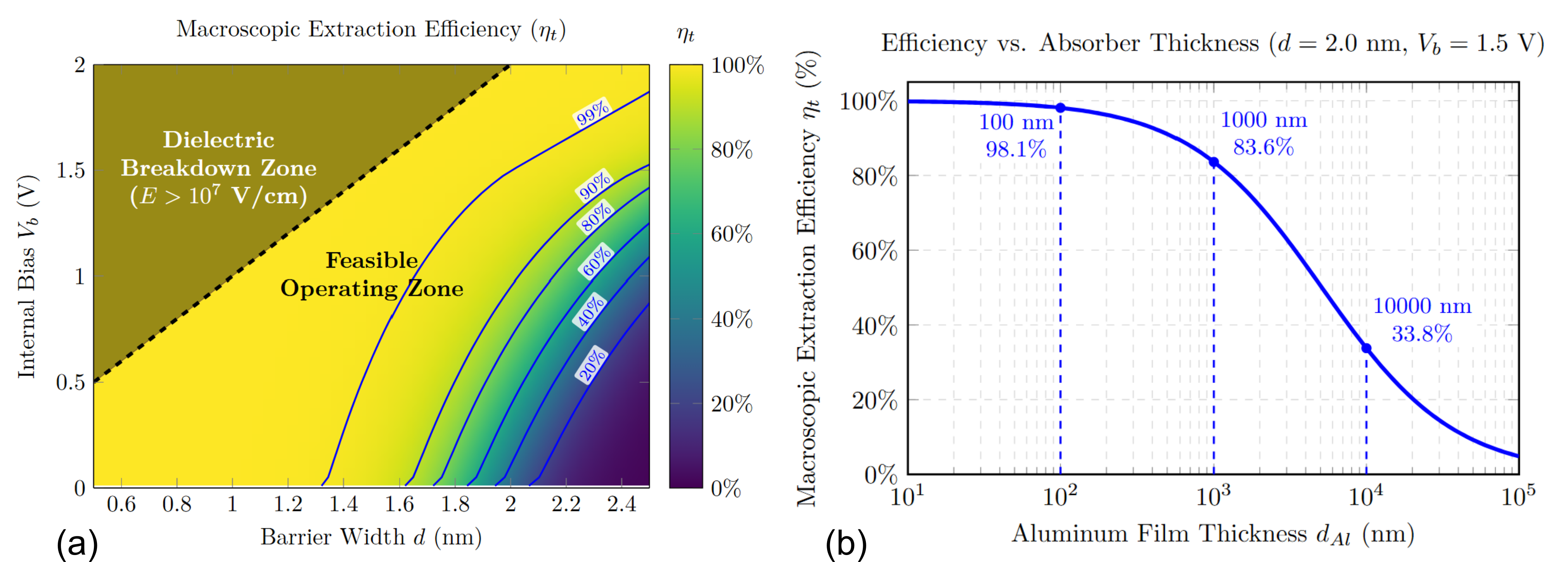}
\caption{Kinetic competition and macroscopic extraction efficiency ($\eta_t$) of quantum tunneling quasiparticles. (a) A two-dimensional thermodynamic surface plot illustrating $\eta_t$ as a function of the insulating barrier width ($d$) and the internal bias voltage ($V_b$). The calculation assumes an optimal superconducting aluminum film thickness of 50~nm, corresponding to a quasiparticle attempt-to-escape frequency of $\nu_0\approx 10^{13}$~Hz. The dashed boundary delineates the safe dielectric breakdown limit ($E\leq 10^7\text{ V/cm}$), establishing a highly tunable feasible operating zone where extraction efficiency can be modulated from 0\% to near 100\%. At the optimal configuration of $d=2.0$~nm and $V_b=1.5$~V, the transmission probability heavily outpaces thermal loss, allowing $\eta_t$ to exceed 99\% efficiency. (b) The macroscopic extraction efficiency evaluated as a function of the aluminum film thickness ($d_{\text{Al}}$) on a logarithmic scale, with barrier parameters fixed at $d=2.0$~nm and $V_b=1.5$~V. Thicker absorber films increase the quasiparticle transit distance, thereby lowering the attempt-to-escape frequency and the resulting extraction efficiency. The curve demonstrates that maintaining the absorber in the thin-film regime ($d_{\text{Al}}<100$~nm) ensures the tunneling process resolves in under 7.8~ns. This nanosecond-scale kinetic transit is overwhelmingly faster than the intrinsic thermal recombination lifetime, consistently conserving over 98\% of the initial signal prior to avalanche multiplication.}
\end{figure}

\subsection{Electron Transit and Drift Simulation}
Electrons successfully completing the quantum tunneling phase are injected into the neutral surface region of the P-type semiconductor layer. Absent a macroscopic external electric field, these minority carriers rely strictly upon slow thermal diffusion to traverse this region, which would inevitably lead to fatal recombination with the majority hole population before reaching the avalanche zone \cite{sze2006}. To eradicate this attenuation and resolve the first fundamental uncertainty of the SuperEM architecture, extensive TCAD signal transport simulations were conducted to optimize the structural doping profile.

Figure 3(a) illustrates the cross-sectional model of the SuperEM semiconductor backend. The device is modeled on a 15~$\mu$m thick, high-resistivity P-type silicon substrate with a baseline doping concentration of $1 \times 10^{12}$~cm$^{-3}$. Beneath the surface interface, a $\sim 2~\mu$m thick P$^+$ gain layer is integrated with a doping concentration of $1 \times 10^{16}$~cm$^{-3}$ to shape the multiplication field. The top surface features a 500~nm thick P$^{++}$ annular ring cathode ($1 \times 10^{20}$~cm$^{-3}$) of diameter $L$, while the substrate is bounded across the entire bottom surface by a 500~nm thick N$^{++}$ anode heavily doped to $1 \times 10^{20}$~cm$^{-3}$. (Note that the superficial superconducting and insulating layers are treated as boundary injection conditions in this model for computational simplicity.)

Upon the application of the 50~V external bias voltage, the internal potential distribution is established. Figures 3(b) and 3(c) quantitatively map the corresponding electric field magnitude along the lateral surface axis (Cutline\_X) and the longitudinal depth axis (Cutline\_Y) intersecting Point A, which is located at the geometric center of the pixel interface. 

As shown in Figure 3(b), the lateral electric field exhibits a distinct minimum at the pixel center and rises sharply toward the peripheral P$^{++}$ annular cathode. This spatial distribution dictates that electrons injected near the center experience a weakened initial lateral driving force, which directly accounts for the geometry-dependent lateral drift delay. Conversely, Figure 3(c) demonstrates that the longitudinal electric field strengthens continuously from the top surface down toward the N$^{++}$ anode bounding the entire bottom surface. This vertical gradient guarantees that once the minority electrons clear the central surface region, they are rapidly accelerated downward to their saturated terminal velocity, culminating in robust avalanche multiplication.

The simulation results therefore dictate two critical architectural mandates:
\begin{itemize}
    \item \textbf{Undoped Interface:} The design must feature a specific undoped region, allowing the silicon to directly contact the tunneling insulator. Simulations confirm that injecting electrons directly into heavily doped zones substantially increases the probability of recombination, weakening the ultimate signal strength.
    \item \textbf{External Drift Electric Field:} To accelerate the tunneled electrons across the interface, a strong external bias voltage is applied across the epitaxial layer. Under the intense electrostatic acceleration provided by this external drift field ($>$500~V/cm), simulations demonstrate that tunneled electrons transition immediately to saturated high-speed vertical drift without deleterious lateral diffusion, negating the need for a complex built-in gradient doping profile.
\end{itemize}

To explicitly quantify the survival probability ($\eta_d$), we model the transit across the 15~$\mu$m high-resistivity substrate, which constitutes the primary recombination zone. Under an externally applied electric field averaging $10^3$~V/cm, acoustic phonon scattering is severely suppressed at deep-cryogenic temperatures, allowing minority electrons to reach a saturated drift velocity ($v_s$) approximating $10^7\text{ cm/s}$ \cite{canali1975}. Utilizing the simulated substrate thickness ($d$) of 15~$\mu$m ($1.5\times 10^{-3}\text{ cm}$), the physical bulk transit time ($t_{\text{drift}}$) evaluates to:
\begin{equation}
t_{\text{drift}} = \frac{d}{v_s} = \frac{1.5\times 10^{-3}\text{ cm}}{10^7\text{ cm/s}} = 150\text{ ps}
\end{equation}
The statistical probability of the electron surviving this bulk transit is governed by the cryogenic minority carrier lifetime ($\tau_n$). Assuming a high-purity silicon substrate where deep-level impurity concentrations are strictly controlled, $\tau_n$ at 10~mK is estimated to be on the order of 1~$\mu$s \cite{brink2006, kelsey2023}, aligning with fundamental solid-state models for high-resistivity substrates \cite{knoll2010}. Consequently, the theoretical bulk survival probability is calculated as:
\begin{equation}
\eta_{\text{bulk}} = \exp\left(-\frac{t_{\text{drift}}}{\tau_n}\right) = \exp\left(-\frac{150\text{ ps}}{1~\mu\text{s}}\right) \approx 0.99985
\end{equation}
However, because our TCAD model treats the upper heterostructure strictly as a boundary injection condition for computational efficiency, this bulk calculation does not natively account for interface trapping at the Al$_2$O$_3$-Si boundary. Factoring in anticipated surface recombination velocities ($S_0$) driven by ALD interface trap densities ($D_{\text{it}}$), the effective total survival probability ($\eta_d$) is projected to systematically drop to approximately 0.95. As demonstrated in Figure 4, this adjusted threshold remains highly compatible with the requirements for sub-meV, high-resolution performance \cite{abazajian2019}.

As demonstrated in Figure 3(d), while the saturated vertical bulk transit resolves in mere picoseconds, the total transient electron drift time is heavily dominated by the initial lateral drift and is therefore strongly dependent on the annular cathode diameter $L$. Because the electric field gradient is weakest at the exact center of the pixel interface (Point A), electrons injected here experience a slower initial drift before reaching the high-field multiplication zone. Consequently, electrons traversing from point A to the anode require approximately 2~ns for a 30~$\mu$m diameter pixel, scaling up to $\sim 25$~ns for a 60~$\mu$m diameter as the central low-field region expands. Consequently, the simulated transient anode current (Figure 3(e)) exhibits a distinct bimodal temporal profile, where the initial sharp peak corresponds to the primary electron drift and avalanche, followed immediately by the secondary drift of the generated holes.

Signal transport simulations conclusively demonstrate that the macroscopic charge collection process—from interface injection to total avalanche multiplication and capacitive signal formation—resolves seamlessly within a 2 to 25~ns transient window, depending strictly on the chosen pixel geometry. Because the overall temporal performance of the detector is determined by the sequential combination of the initial quantum tunneling extraction ($\tau_{\text{tun}}\leq 7.8$~ns) and this subsequent avalanche transit, the total macroscopic signal completely resolves in under 35~ns. The fundamental physical reason SuperEM responds orders of magnitude faster than traditional calorimeters is that it extracts charge carriers directly from the non-equilibrium quasiparticle state immediately following the pair-breaking event. In contrast, traditional Transition Edge Sensors (TES) and similar architectures must wait for complete thermal relaxation to measure the superconducting quenching effect caused by the macroscopic temperature rise of the equilibrium state. This fast, kinetically driven response time strictly avoids the macroscopic signal pile-up that plagues thermodynamic sensors, proving the high-efficiency transport mechanics of the SuperEM design.

\begin{figure}[H]
\centering
\includegraphics[width=0.9\textwidth]{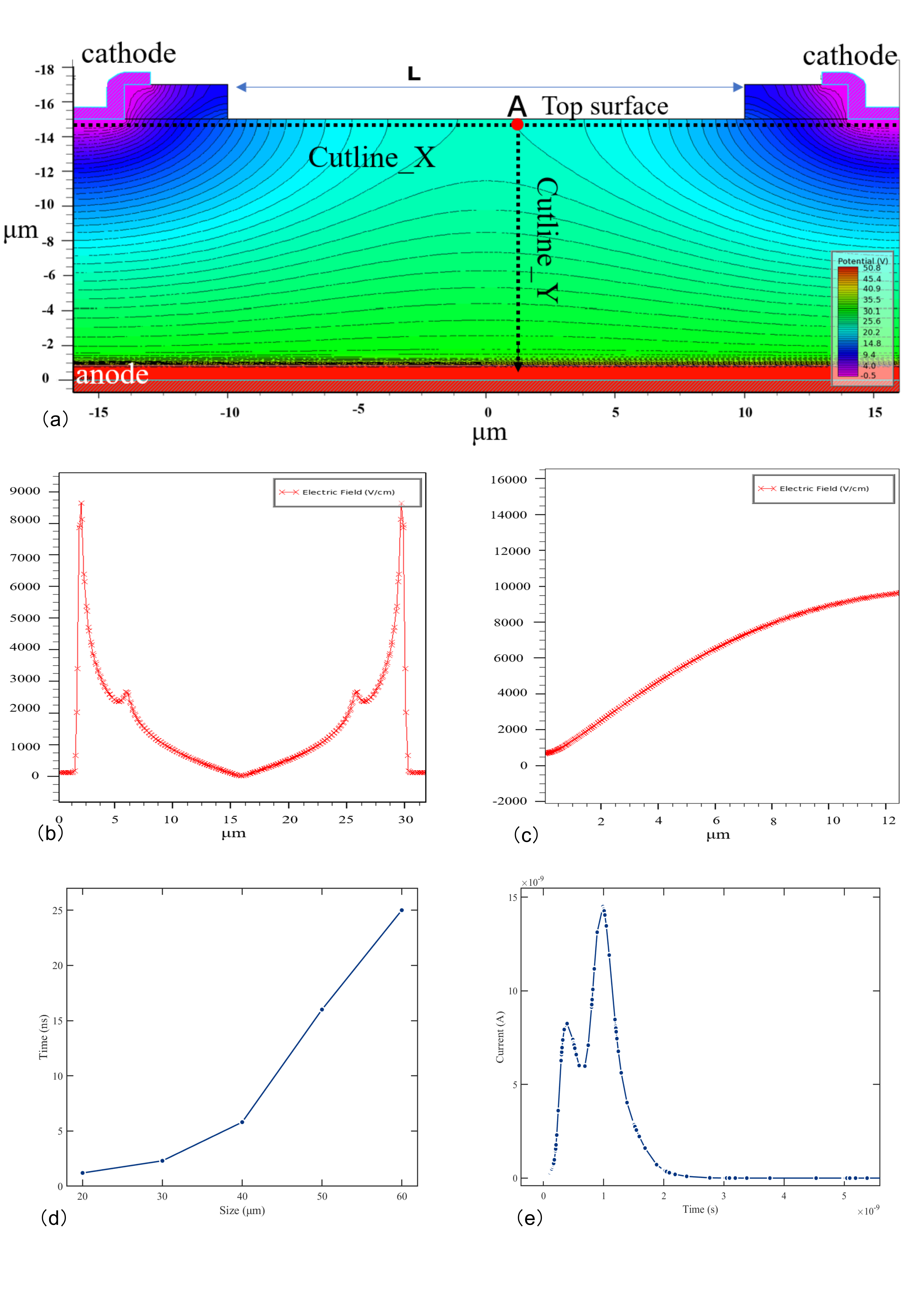}
\caption{TCAD signal transport and transient response simulation. (a) 2D cross-sectional TCAD model demonstrating the internal potential distribution under an external bias. (b) Simulated electric field profile along the lateral surface axis (Cutline\_X). (c) Electric field profile along the longitudinal depth axis (Cutline\_Y). (d) Electron drift time as a function of the annular cathode diameter $L$, ranging from 2~ns for 30~$\mu$m to 25~ns for 60~$\mu$m pixels. (e) Simulated transient anode current exhibiting the primary electron and secondary hole drift peaks, resolving completely within nanoseconds.}
\end{figure}

\subsection{Cryogenic Multiplication Region Dynamics}
As mapped in the physical signal pathway flowchart of Figure 4(a), the ultimate absolute energy resolution of the detector is determined by the progressive accumulation of statistical variance. This is mathematically formalized using an Effective Fano Factor ($F_{eff}$) derived via the Burgess Variance Theorem for cascaded events \cite{knoll2010}. Integrating the binomial partition noise of the independent survival probabilities and the relative multiplication variance, the generalized equation is expressed as:  

\begin{equation}
F_{eff} = F + \frac{1 - \eta_t \eta_d}{\eta_t \eta_d} + \frac{F_a - 1}{\eta_t \eta_d}
\end{equation}
where each parameter quantifies the statistical variance introduced at a distinct stage:  

\begin{itemize}
    \item \textbf{$F$ (Intrinsic Fano Factor):} Represents the fundamental statistical fluctuation in the initial number of Cooper pairs broken within the superconducting absorber ($F \approx 0.2$).  
    \item \textbf{$\eta_t$ (Tunneling Extraction Probability) and $\eta_d$ (Drift Survival Probability):} The combined term $\frac{1 - \eta_t \eta_d}{\eta_t \eta_d}$ accounts for the binomial partition noise introduced by quasiparticles failing to successfully cross the quantum tunneling barrier, or minority electrons recombining during their transit across the neutral epitaxial layer.  
    \item \textbf{$F_a$ (Excess Noise Factor):} Encapsulates the stochastic variance introduced by the history of random impact ionization collisions during the macroscopic avalanche multiplication process.
\end{itemize}

If the detector operates in the Monolithic Proportional Mode, linear impact ionization is dictated by the McIntyre multiplication noise theory \cite{sze2006}. For an electron-injected avalanche in a silicon lattice, the Excess Noise Factor ($F_a$) is strictly bounded near a theoretical minimum limit of $F_a \approx 2.0$ \cite{sze2006}. Assuming highly optimized physical configurations ($\eta_t = 0.8$, $\eta_d = 0.95$) and this best-case linear scenario ($F_a = 2.0$), the Effective Fano Factor evaluates to $F_{eff} \approx 1.83$. Recalculating the absolute energy resolution (FWHM) at the 1~eV milestone using this proportional variance penalty yields approximately 77.7~meV.

This operational mode fails to meet the strict $\le 40$~meV resolution target, which is mathematically dictated by the $\sim 50$~meV minimum effective mass bound of the inverted hierarchy, preventing the physical differentiation of neutrino mass orderings \cite{aker2019, ashtari2017}. Figure 4(b) visually confirms this severe limitation. Because the theoretical minimum resolution for linear multiplication is mathematically floored at $\sim 62.9$~meV (assuming a perfect, 100\% signal efficiency where $\eta_t = 1.0$ and $\eta_d = 1.0$), it is physically impossible to plot a 50~meV contour for the Monolithic Proportional Mode. Instead, the red curve in Figure 4(b) represents an 80~meV boundary, demonstrating that even with near-perfect quantum extraction and drift survival, proportional amplification remains fundamentally restricted from reaching the precision required for next-generation neutrino physics.  

Consequently, the general monolithic S-I-P-N architecture is insufficient for precision neutrino physics. The transition to a highly dense, Pixelated Geiger-Counting Mode across the detector plane arrays is not merely an operational alternative, but the fundamental architectural novelty required for ultimate resolution. By dividing the active area into micro-pixels, the total macroscopic charge released ($Q_p = C_{pix} \Delta V$, where $C_{pix}$ is the micro-pixel capacitance and $\Delta V$ is the applied overvoltage) is decoupled from the stochastic history of impact ionization collisions. This topology physically truncates the mathematical variance of the amplification gain to strictly zero ($\sigma_M^2 \rightarrow 0$), transitioning the macroscopic excess noise factor to $F_a = 1 + P_{CT} + P_{AP}$ \cite{acerbi2019}.  

Because deep-level thermal detrapping is entirely frozen out at 10~mK, afterpulsing is mathematically nullified ($P_{AP} \approx 0$). Consequently, the multiplication penalty simplifies exclusively to the prompt optical crosstalk probability ($F_a - 1 \approx P_{CT}$). Applying the combined signal efficiency ($\eta_t \eta_d = 0.76$) and a conservative 5\% optical crosstalk penalty ($P_{CT} = 0.05$) typical of deep-trench isolated arrays, the Effective Fano Factor collapses to $F_{eff} \approx 0.582$. With this restricted variance, the absolute intrinsic energy resolution at 1~eV definitively reaches $\sim 43.8$~meV.

While this marginally exceeds the strict $\leq 40$~meV instrumental target widely cited for isolating the inverted mass hierarchy, it physically encroaches upon the required boundary. Future optimizations in pixel geometry and minimizing systematic variances in the internal bias voltage ($V_b$) will be imperative to compress this resolution to the sub-40~meV regime. Nonetheless, Figure 4(b) provides a comprehensive visualization of this architectural advantage. The underlying thermogram heatmap maps the absolute energy resolution for the Pixelated Geiger-Counting Mode (incorporating the $F_a = 1.05$ crosstalk penalty). The solid white contour delineates the stringent 50~meV resolution boundary. As geometrically illustrated by the expansive parameter space above and to the right of this white curve, the Geiger mode provides a highly forgiving operational window. It mathematically guarantees that even if non-ideal fabrication defects cause the tunneling extraction efficiency ($\eta_t$) or the drift survival probability ($\eta_d$) to independently drop to roughly 80\%, the detector will still robustly isolate the signal within the required high-resolution regime.

While the Pixelated Geiger-Counting Mode is theoretically optimal for minimizing variance, its practical implementation requires rigorous management of the detector’s dynamic range and linearity. Based on the 0.595~meV effective excitation energy, a 1~eV energy deposition generates approximately 1,680 primary quasiparticles. For the detector to maintain strict energy proportionality, each extracted electron must trigger an independent, inactive micro-pixel. If multiple charge carriers tunnel into the exact same micro-pixel simultaneously, they will trigger only a single saturated avalanche. This phenomenon, analogous to signal pile-up, inherently destroys charge quantization and severely degrades the absolute energy resolution. To sustain a highly linear response up to the 1~eV scale with a negligible saturation probability, the semiconductor backend must possess an ultra-high pixel density, typically on the order of $10^4$ to $10^5$ micro-pixels per mm$^2$ (corresponding to microscopic pixel pitches of roughly 3 to 10~$\mu$m). Furthermore, maximizing the geometric fill factor is imperative to ensure that non-equilibrium quasiparticles do not tunnel into inactive dead zones between the avalanche regions. Because traditional 2D surface routing of quenching resistors and trenches severely degrades the fill factor at microscopic pixel pitches, achieving this dynamic range will likely necessitate advanced 3D vertical integration techniques (such as Through-Silicon Vias) to bury inactive components beneath the avalanche region, ensuring the preservation of both sub-meV precision and macroscopic linearity.

\begin{figure}[H]
\centering
\includegraphics[width=0.9\textwidth]{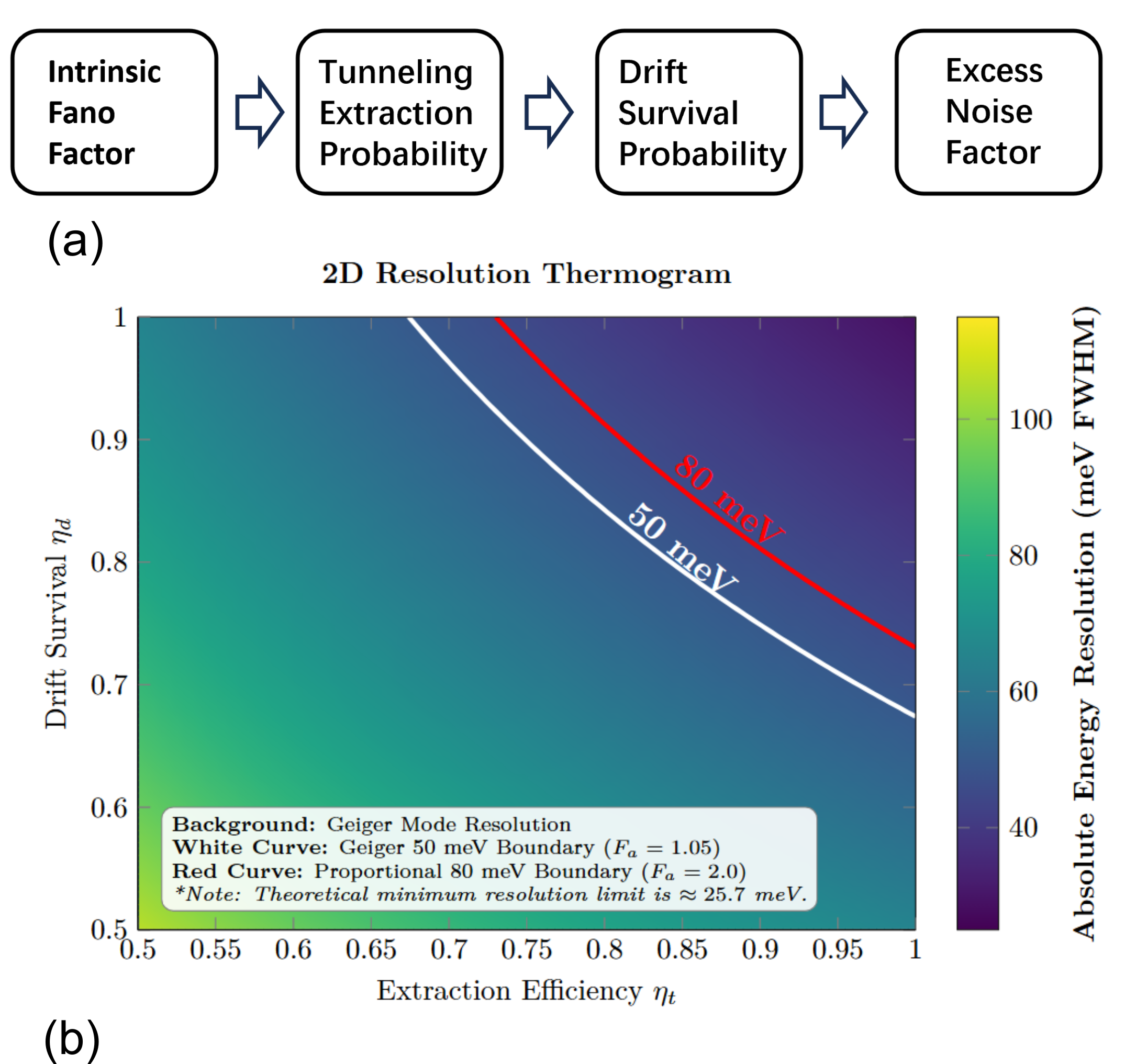}
\caption{Theoretical energy resolution and statistical variance propagation. (a) Flowchart mapping the physical signal pathway from initial quantum absorption to macroscopic avalanche amplification, detailing the progressive accumulation of statistical variance as governed by the Burgess Variance Theorem. (b) A two-dimensional thermogram mapping the absolute energy resolution (FWHM) for a 1~eV energy deposition as a function of the tunneling extraction efficiency ($\eta_t$) and drift survival probability ($\eta_d$). The underlying heatmap displays the high-resolution parameter space achieved by the Pixelated Geiger-Counting Mode, integrating a conservative 5\% optical crosstalk penalty ($F_a = 1.05$). The solid contour lines delineate key resolution boundaries: the stringent 50~meV target for the Geiger-Counting mode (white curve) and the 80~meV contour for the Monolithic Proportional Mode (red curve, strictly bounded by the ideal linear limit of $F_a = 2.0$). The plot geometrically demonstrates that the Geiger mode provides a vastly expanded, highly forgiving operational parameter space. In contrast, the theoretical minimum resolution of the Proportional mode is mathematically restricted to $\sim 62.9$~meV, physically barring it from reaching the sub-50~meV targets required for precision neutrino mass experiments regardless of interface efficiency.}
\end{figure}

\section{Experimental Validation of Avalanche Dynamics at 10 mK}
With the drift mechanics verified via simulation, the viability of the SuperEM concept entirely depends upon the functional integrity of the silicon amplification region at ultra-low temperatures. At $\sim$10~mK, severe carrier freeze-out threatens to stifle the avalanche process by eliminating free carriers and fundamentally altering electric field distributions \cite{sze2006}.

As the first empirical milestone of this detector development program, our recent component-level experimental work (currently available as a preprint \cite{gao2025}) conducted comprehensive experiments within a dilution refrigerator, utilizing commercial Silicon Photomultiplier (SiPM) arrays as a proxy to characterize silicon PN junction behavior at a stabilized base temperature of 10~mK. These empirical results definitively confirm that silicon devices retain robust, quantized avalanche capabilities in the deep cryogenic regime.

The characterization elucidated a distinct interplay of competing physical mechanisms governing the breakdown voltage ($V_{\text{bd}}$). Although $V_{\text{bd}}$ initially decreases from 293~K to 50~K in response to the suppression of phonon scattering, it exhibits an anomalous increase as the temperature is further reduced to 10~mK. This behavior serves as a direct manifestation of carrier freeze-out, which necessitates the application of a heightened electric field to instigate the avalanche process \cite{gao2025}. Furthermore, at 10~mK, lattice scattering is effectively eradicated; this extends the carrier mean free path and precipitates a pronounced elevation in the impact ionization coefficients ($\alpha, \beta$) \cite{gao2025, sze2006}.

Despite these altered physical dynamics, the quantized nature of the avalanche process is strictly preserved. At an applied overvoltage of 2.5~V—operating in the Geiger-counting regime—the silicon substrate exhibited a highly stable intrinsic single-carrier gain on the order of $10^6$ \cite{gao2025}. Significantly, operation at the 10~mK base temperature facilitates a substantial attenuation of thermal noise. Empirical measurements recorded a primary dark count rate (DCR) of 5~mHz/mm$^2$. Rather than merely representing a seven-order-of-magnitude decrease relative to standard room-temperature operation \cite{gao2025}, this exceptionally low dark rate proves that deeply cooled semiconductor amplification stages are highly competitive with the thermal background profiles of existing cryogenic calorimeters (e.g., TES and KID arrays). While these commercial SiPM proxy experiments conclusively validate the survival of the impact ionization mechanism at 10~mK, it must be noted that their standard doping profiles do not replicate the specialized undoped interface modeled in our TCAD simulations. Therefore, the integrated extraction efficiency ($\eta_t$) and drift survival probability ($\eta_d$) remain subject to systematic errors, including localized trap-assisted tunneling (TAT) from ALD defects and RC delays introduced by device capacitance, which must be quantified in a future monolithic prototype.

While the semiconductor amplification stage proves exceptionally quiet at 10~mK, the overarching dark rate of the fully integrated S-I-P-N device will also be dictated by spontaneous quasiparticle generation within the superconducting Aluminum absorber. Analogous to the ``quasiparticle poisoning'' that severely degrades coherence times in superconducting qubits \cite{li2025}, the S-layer is highly susceptible to trace energy depositions from ambient radioactivity, cosmic ray muons, and stray THz or far-infrared photons propagating through the cryostat shielding. Because the Cooper pair binding energy is exceedingly low ($\sim$0.35~meV), even an isolated infrared photon can break pairs and generate a localized low-level quasiparticle population, threatening to trigger false-positive avalanches. Fortunately, the SuperEM architecture provides a native, robust defense against this background via the tunable internal bias ($V_b$) applied across the ALD insulating layer. By dynamically calibrating $V_b$, the quantum tunneling potential barrier acts as an active electronic discriminator. Low-density quasiparticle fluctuations induced by stray photons or trace radioactivity can be effectively thresholded out by maintaining $V_b$ just below the critical Quasi-Fowler-Nordheim transmission regime. Genuine rare-event signal interactions, however, generate a massive, instantaneous localized burst of quasiparticles (e.g., thousands of carriers for a 1~eV deposition) that collectively overcome this finely tuned barrier. This tunable gating effectively shields the high-gain semiconductor from environmental quasiparticle poisoning while preserving unparalleled sensitivity to sub-meV scale physical signals.

While engineering challenges remain—specifically managing the ``second divergence'' caused by correlated noises (such as afterpulsing and optical crosstalk), which currently limits the stable operational overvoltage window to approximately 5~V \cite{gao2025}—this successful empirical verification of 10~mK quantized avalanche generation completely clears the primary physical obstacle regarding the semiconductor backend. Therefore, with the macroscopic amplification stage experimentally validated, the remaining challenge is purely the integration of the quantum tunneling interface.

\section{Conclusions}
By theoretically marrying the sub-meV excitation thresholds of superconducting absorbers with the ultra-low-noise, high-gain digitization of pixelated Geiger-mode semiconductors, the original SuperEM S-I-P-N architecture provides a highly viable structural pathway to bypassing the 'impossible triangle' of rare-event detection. The resolution of the primary physical uncertainties presented herein transitions this detector concept into a robust framework for forthcoming experimental realization.

Our drift simulations confirm that optimized undoped interfaces and a strong external drift electric field permit rapid, nanosecond-scale electron transit (resolving completely in under 35~ns) while maintaining collection efficiencies exceeding 95\%. Simultaneously, our prior deep-cryogenic experiments prove that silicon avalanche junctions maintain robust gain and near-zero thermal noise at 10~mK \cite{gao2025}. While commercial standard-doping profiles served as successful proxies for this macroscopic amplification stage, validating that the requisite undoped epitaxial layer can identically sustain these high-field gradients without premature breakdown remains the primary objective for upcoming monolithic prototypes.

Despite the profound theoretical and empirical validations of the SuperEM architecture, transitioning this fundamental device innovation into a macroscopic, scalable observatory presents unprecedented engineering challenges. First, the fabrication of the S-I-P-N heterostructure relies upon the extreme precision of the quantum tunneling barrier. Achieving a highly uniform, defect-free Al$_2$O$_3$ insulating film precisely at the 1.0 to 3.0~nm scale across large-area wafers is a formidable materials science challenge. Any nanoscale pinhole defects, lattice mismatches, or local thickness non-uniformities introduced during the Atomic Layer Deposition (ALD) process would exponentially amplify parasitic leakage currents, fundamentally compromising the selectivity of the Quasi-Fowler-Nordheim tunneling mechanism.

Furthermore, while the pixelated Geiger-mode guarantees single-electron quantization, scaling to large-area macroscopic arrays introduces a severe cryogenic readout challenge. The dilution refrigerators required to maintain the 10~mK operational environment possess highly restrictive cooling power budgets, typically limited to microwatts at the mixing chamber. Placing active amplification and digitization electronics directly adjacent to the detector array is thermodynamically unfeasible. Consequently, future deployment necessitates the development of ultra-low-power cryogenic Application-Specific Integrated Circuits (cryo-CMOS ASICs) operating at the higher 4~K thermal stage. Bridging the 10~mK sensors to these 4~K ASICs via high-bandwidth, low-thermal-conductivity transmission lines will be required to process the nanosecond transient avalanches without exceeding the power limit of the refrigerator.

With the fundamental viability of the individual avalanche components \cite{gao2025} and the overarching S-I-P-N drift architecture established herein, our research program is actively transitioning into its next phase: the fabrication, integration, and direct cryogenic characterization of a fully monolithic SuperEM prototype. The mathematical frameworks established in this work demonstrate that bypassing proportional multiplication limits via the Pixelated Geiger-Counting Mode is structurally mandatory; it collapses the macroscopic variance penalty and drives the absolute intrinsic energy resolution to $\sim 43.8$~meV for a 1~eV deposition. For the pursuit of the absolute neutrino mass, crossing the 40~meV energy resolution threshold signifies that hardware can finally match the precision required to resolve the normal versus inverted mass ordering \cite{aker2019, ashtari2017}. In the domain of dark matter detection, circumventing the restrictive 1.12~eV bandgap limit of traditional silicon devices allows the SuperEM detector to scan the entirely uncharted sub-GeV mass parameter space \cite{hochberg2016}. Crucially, for the newly proposed parametric fluorescence method of C$\nu$B detection \cite{huang2026}, the SuperEM detector provides the exact $\mathcal{O}(10)$~meV single-photon sensitivity and scaling to macroscopic target volumes capability.

Beyond fundamental particle astrophysics, the architecture holds transformative utility for applied quantum technologies and observational cosmology. In superconducting quantum computing systems \cite{li2025}, its nanosecond temporal resolution and sub-meV thresholds provide a critical mechanism for the real-time identification and active error correction of quasiparticle poisoning events. Simultaneously, for the next generation of primordial gravitational wave observatories mapping CMB B-mode polarization—such as CMB-S4 \cite{abazajian2019} and AliCPT \cite{li2017}—the architecture directly enables the deployment of massively detector plane arrays capable of unprecedented pixel scale.

\section*{Acknowledgements}
We would like to express our gratitude to Min Yu, Mei Zhao, Qi Yan for their discussion on semiconductor structure, Yifei Zhang and Zhengwei Li for their discussion on superconducting detectors, Yuekun Heng for his discussion on the measurement of the absolute mass of neutrinos, Shun Zhou for his discussion on the detection of the cosmic neutrino background, Junhua Wang and Xuegang Li for their discussion on error correction in superconducting quantum computing, Jingbo Ye for his discussion on readout electronics, and Liangjian Wen and Jun Cao for their discussion on research strategies.

\end{document}